\documentclass[dvips]{article}

\usepackage{icrc2011}

\title{MAGIC detection of VHE $\gamma$-ray emission from NGC~1275 and IC~310}
\newcommand{\etal}{\MakeLowercase{\textit{et al. }}} 
\shorttitle{Hildebrand \etal MAGIC detection of VHE emission from NGC~1275 and IC~310}

\authors{
         Dorothee Hildebrand$^{1}$, 
         Saverio Lombardi$^{2}$, 
         Pierre Colin$^{3}$, 
         Julian Sitarek$^{4}$, 
         Fabio Zandanel$^{5}$ 
         Francisco Prada$^{5}$ 
         for the MAGIC Collaboration, 
         and Christoph Pfrommer$^{6}$, 
         Anders Pinzke$^{7}$}
\afiliations{$^1$ETH Zurich, 8093 Zurich, Switzerland\\
             $^2$Universit\`{a} di Padova and INFN, I-35131 Padova, Italy\\             
             $^3$Max-Planck-Institut f\"{u}r Physik, D-80805 M\"{u}nchen, Germany\\
             $^4$University of {\L}\'{o}d\'{z}, Poland\\              
             $^5$Instituto de Astrof\'{i}sica de Andaluc\'{i}a (IAA-CSIC), E-18080 Granada, Spain\\             
             $^6$Heidelberg Institute for Theoretical Studies (HITS), D-69118 Heidelberg, Germany\\
             $^7$UCSB, California, USA\\
             }
\email{dorothee.hildebrand@phys.ethz.ch}

\abstract{The MAGIC Cherenkov telescopes observed the Perseus cluster sky region in stereo mode 
for nearly 90~hr from October 2009 to February 2011. This campaign led to the discovery of 
very high energy $\gamma$-ray emission from the central radio galaxy NGC~1275 and the head-tail radio
galaxy IC~310. Here we report the results on the most recent discovery of NGC~1275 which was 
detected at low energies in the 2010/2011 data. We also present latest results on IC~310, which had 
been detected in the 2009/2010 data.}

\begin{document}
\maketitle

\section{Introduction}
Most of the previously known $\sim$45 extragalactic very high energy (VHE) $\gamma$-ray emitters 
are blazars, the few exceptions being two radio galaxies, M~87 \cite{lab1} and Cen~A \cite{lab2}, 
and two starburst galaxies, NGC~253 \cite{lab3} and M82 \cite{lab4}. The two galaxies NGC~1275 
and IC~310 recently discovered by the MAGIC telescopes in the Perseus cluster (redshift z~=~0.018) 
do not easily fit to any of the known VHE emitter classes.\\
NGC~1275 is the central galaxy of the Perseus cluster and its classification varies between different 
papers and catalogues. It was included in the original Seyfert catalogue \cite{lab5}, but already 
flagged to be unusual because of its complex structure. After the introduction of Seyfert 1 and 2 
subclasses by Khachikian and Weedman \cite{lab6}, NGC~1275 was labeled as Sy2. Few years later, 
Veron proposed it to be a BL~Lac \cite{lab7}, but in his latest catalogue NGC~1275 is classified as 
Sy1.5 \cite{lab8}. The complex structure of NGC~1275 including surrounding filaments leads to a 
peculiar morphology classification \cite{lab9}. Additionally, another galaxy called High Velocity 
System is moving towards NGC~1275 along the line of sight. Recent observations show that a collision 
between the two galaxies has not started yet \cite{lab10}.\\
IC~310 is classified as head tail radio galaxy, a type of active galactic nuclei only occuring 
in dense galaxy clusters like the Perseus one. Due to the fast movement of IC~310 relative to the 
cluster, the friction between the jet with the intra-cluster medium (ICM) causes a strong bending 
of the jets \cite{lab11, lab12}. In 1999 it was suggested that IC~310 could be a dim blazar because 
of the absence of strong emission lines and the spectral indices on radio and X-ray measurements 
\cite{ic310_blazar}. Later on it was also shown that the X-ray emission may originate from the 
central active galactic nucleus of a BL~Lac-type object \cite{ic310_xray}.\\
The \emph{Fermi}--LAT measured high energy $\gamma$-rays from NGC~1275 up to 25~GeV \cite{lab14} and 
detected strong flaring activity in July 2010 \cite{lab15}. Additionally, an analysis of Fermi data 
resulted in a detection of IC~310 for energies above 30~GeV \cite{lab16}.
%
%
\section{MAGIC Observations}
The MAGIC (Major Atmospheric Gamma Imaging Cherenkov) experiment consists of two 17~meter Imaging Air 
Cherenkov telescopes located at the Canary Island of La Palma, at 2200~meters a.s.l., working in stereoscopic 
mode since the autumn 2009. The MAGIC telescopes are currently the largest world-wide existing Imaging 
Atmospheric Cherenkov Telescopes (IACTs), leading to a low energy threshold of $\sim$50~GeV. Depending on 
the energy of the primary particle, the system has an energy resolution of 15-25$\%$ and a angular resolution 
of 0.05-0.12$^{\circ}$. The telescopes have a field of view of 3.5$^{\circ}$ and can turn to any position 
faster than 40~s. The sensitivity of the MAGIC telescopes is $\sim$0.8$\%$ of the Crab Nebula flux above 
$\sim$250~GeV in 50h of observation time \cite{sitarek2011}.\\
The MAGIC experiment observed the central part of the Perseus cluster for $\sim$25~hr in 2008, when 
only the first telescope was operating (mono data). This survey resulted in upper limits 
on VHE $\gamma$-ray emission above 100~GeV from NGC~1275, intra-cluster cosmic rays and 
dark matter annihilations \cite{lab13}.\\
The stereo observation campaign on NGC~1275 and on the central part of the Perseus cluster took place from 
October 2009 to February 2011, during dark nights and observing the source at the lowest possible zenith 
angles, which guaranteed a low energy threshold. The data can be divided in two main samples, corresponding 
respectively to the observational seasons from October 2009 to February 2010, and from August 2010 to February 2011. 
During the first season, $\sim$45~hr of observation time were accumulated using two wobble \cite{fomin1994}
positions offset by 0.4$^\circ$ in RA with respect to NGC~1275. Since the commissioning of the stereo trigger 
system was still ongoing, the observations were carried out in the so-called \emph{soft stereo mode}, i.e. 
using the MAGIC-I trigger system and reading out the second telescope simultaneously. This results in slightly higher threshold, but allows a partially independent mono analysis. Conversely, for the 
second season, the full stereo trigger and four wobble pointing positions equally distributed around NGC~1275 
were used. With this settings, a higher sensitivity and a better coverage of the central 
part of the Perseus cluster were achieved. Additional $\sim$45~hr of observation time were added to the 
previous 2009/2010 dataset.\\
%
%

%
\section{Data Analysis and Results}
\begin{figure}[hbt!]
\vspace{5mm}
\centering
\includegraphics[width=3.in]{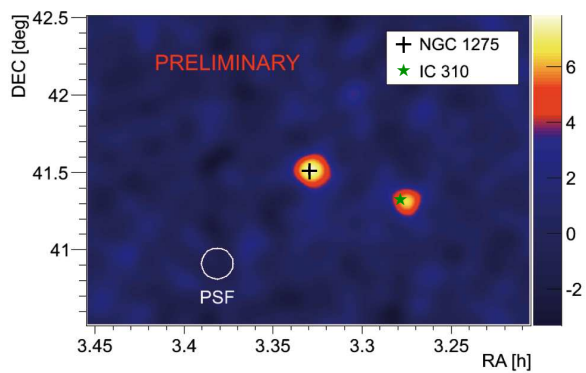}
\caption{
Significance skymap above 150~GeV of the Perseus cluster region. 
For this map the overall MAGIC stereo data from October 2009 to February 2011 have been combined. 
NGC~1275 and IC~310 are clearly detected by MAGIC.
}
\label{fig1}
\end{figure}

The analysis of the data was performed with the standard MAGIC reconstruction and analysis software \cite{lombardi2011a}. 
Due to the different trigger conditions used in the two seasons, the datasets had to be analyzed seperately. The overall 
stereo campaign resulted in the detection at VHE of both NGC~1275 and IC~310, as shown in the significance skymap above 
150~GeV in figure~\ref{fig1}. In the following, we report the main MAGIC results achieved so far on these two sources. 
%
%
\subsection{IC~310}
\begin{figure}[hbt!]
\vspace{5mm}
\centering
\includegraphics[width=3.in]{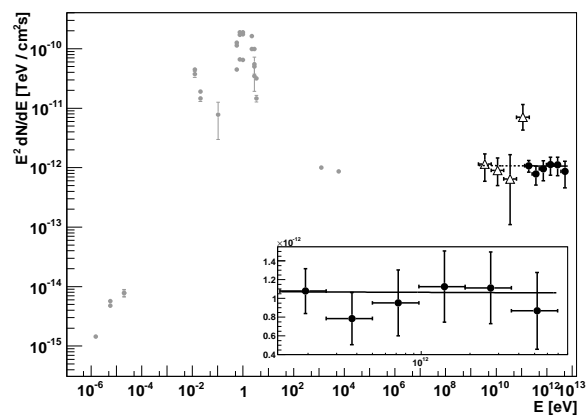}
\caption{
Spectral energy distribution of IC~310 obtained with 20.6~hr of MAGIC stereo data (full circles). 
Open triangles show the flux measurements from the \emph{Fermi}--LAT from its first 
two years of operation. Archival X-ray, optical, IR and radio data obtained from 
the NED database are shown with dots. The solid line shows a power-law fit 
to the MAGIC data, and the dotted line is its extrapolation to the GeV energies.
A zoom-in of the MAGIC points is also shown \cite{ic310_letter}.
}
\label{fig2}
\end{figure}

\begin{figure}[hbt!]
\vspace{5mm}
\centering
\includegraphics[width=3.in]{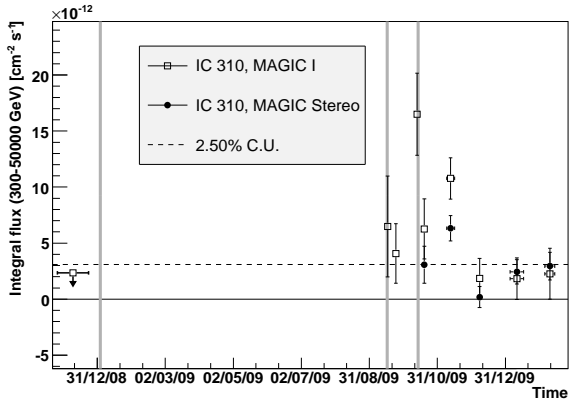}
\caption{
Lightcurve (in 10~day bins) of the $\gamma$-ray emission of IC~310 above 300~GeV obtained with 
the mono (open squares) and the stereo (full circles) MAGIC data from 2009/2010 campaign. 
The open square with an arrow is the upper limit on the emission achieved from mono data 
taken between November and December 2008. Vertical gray lines show the arrival times 
of $>$100 GeV photons from the \emph{Fermi}--LAT instrument. The horizontal dashed line 
is a flux level of 2.5$\%$ Crab Units (C.U.) \cite{ic310_letter}.}
\label{fig3}
\end{figure}

\begin{figure}[hbt!]
\vspace{5mm}
\centering
\includegraphics[width=3.in]{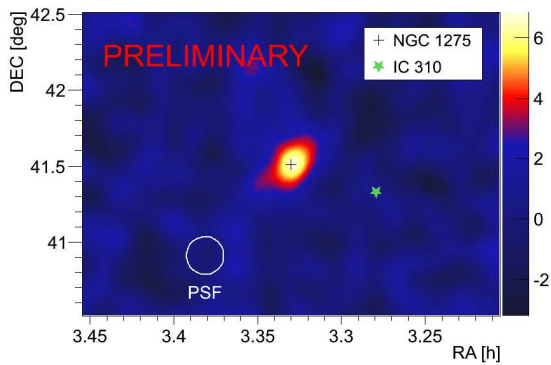}
\caption{
Significance skymap above 100~GeV of the Perseus cluster region. 
For this map the MAGIC stereo data from August 2010 to February 2011 have been used. 
NGC~1275 is clearly detected at the center of the cluster, while IC~310 is not visible 
at VHE.
}
\label{fig4}
\end{figure}

IC~310 was serendipitously discovered in the field of view of the NGC~1275 observation during the 2009/2010 survey.
Due to angular acceptance reason, only one of the two wobble pointing position datasets was used for the detection and
for the flux estimation of the source, resulting in $\sim$20~hr of observation time. 
The source was detected both in mono and stereo mode, and it was the first source discovered by MAGIC using stereoscopy.\\
The spectrum obtained from these data combined with results from Fermi shows an unusual hard spectral index 
$\Gamma\sim$-2.00 in an energy range between 2~GeV and 7~TeV, without any hint of a cut-off (see figure~\ref{fig2}). 
The lightcurve obtained from 2008 to 2010 mono and stereo datasets shows a strong hint of variability, 
as reported in figure~\ref{fig3}. A preliminary analysis of the 2010/2011 data confirms the variability 
of the source (on year scale) since the object shows significantly weaker VHE emissions in this period. 
This is displayed in figure~\ref{fig4}, where the significance skymap above 100~GeV (from data taken 
between August 2010 and February 2011) shows a point-like emission from NGC~1275, whereas no hint of 
signal coming from IC~310 is present. The estimation from the 2010/2011 dataset of the flux and the 
lightcurve of IC~310 is still ongoing. 
%
%
%
\subsection{NGC~1275}
\begin{figure}[hbt!]
\vspace{5mm}
\centering
\includegraphics[width=3.in]{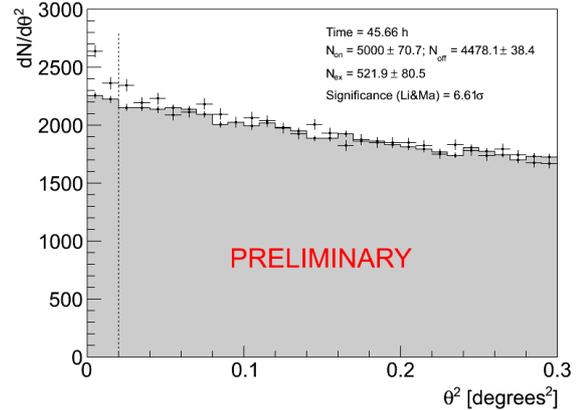}
\caption{
MAGIC stereo $\theta^2$ plot of NGC~1275 above 100~GeV from 2010/2011 stereo data. The black points represent the 
\emph{signal} while the gray shaded region is the \emph{background}. The vertical dotted line defines 
the \emph{signal region} below which we integrate the signal and calculate the backround. A clear signal at $\sim$6.6$\sigma$
level is detected.
}
\label{fig5}
\end{figure}

The 2009/2010 data analysis showed a hint of VHE emission from NGC~1275. The data taken was continued 
in August 2010 which finally resulted in the detection of the source in October 2010, announced in the ATel\#2916. 
In figure~\ref{fig5}, the $\theta^{2}$ plot above 100~GeV from the 2010/2011 survey (after data selection) is shown:
a clear $\gamma$-ray emission at $\sim$6.6$\sigma$ level is detected (see also figure\ref{fig4}). The signal decreases rapidly 
with energy and vanishes above approximately 600~GeV \cite{lombardi2011b}.

The unfolded spectrum of NGC1275 is shown in figure \ref{fig4}. It can be fitted with a power law with a very soft index of about -\,4. 

\begin{figure}[hbt!]
\vspace{5mm}
\centering
\includegraphics[width=3.in]{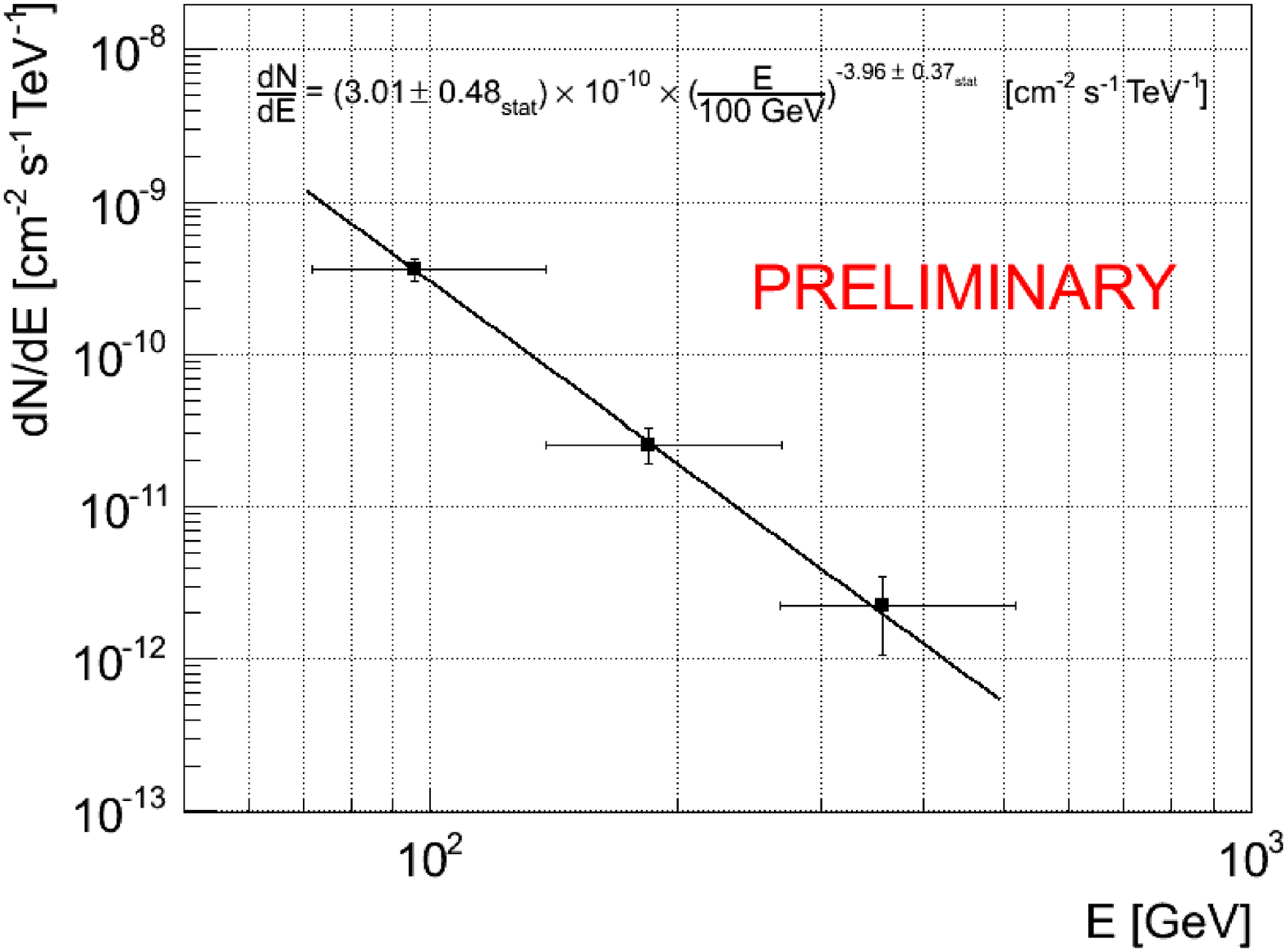}
\caption{
MAGIC unfolded stereo spectrum of NGC~1275 using data from 2010/2011. The spectrum is fitted by a power law with very soft index of about -\,4. Due to the unfolding process the points and errors are partially correlated.
}
\label{fig6}
\end{figure}

%
\section{Discussion}
We have reported the current results on the galaxies IC~310 and NGC~1275 achieved from the MAGIC stereo observations
of the Perseus galaxy cluster region. The two sources have been detected respectively in 2009/2010 and 2010/2011 
dataset. \\
With IC~310 being significantly weaker in the season 2010/11 (see Figure~\ref{fig4}) than in the previous 
season, the variability of the source is proven. This excludes the emission mechanism via bow shock model 
as proposed by Neronov \etal \cite{lab16}, and favours the emission mechanism to be BL~Lac like, where a part of 
the inner jets points towards Earth.\\
The detection of NGC~1275 was achieved thanks to excellent sensitivity of the MAGIC telescopes around $\sim$100~GeV,
which makes MAGIC the leading ground-based instrument currently operating at those energies. A detailed estimation of 
the spectrum and lightcurve of the source is still ongoing and will give the possibility to test models of active galactic nuclei. 
Additionally, as NGC~1275 is the main foreground for intracluster cosmic ray and dark matter searches, a detailed
knowledge of the spectrum allows a better understanding of the cluster itself. A detailed description of the Perseus 
cluster and its physics can be found in \cite{lombardi2011b}. \\
%
%
\section{Acknowledgements}
We would like to thank the Instituto de Astrof\'{\i}sica de
Canarias for the excellent working conditions at the
Observatorio del Roque de los Muchachos in La Palma.
The support of the German BMBF and MPG, the Italian INFN, 
the Swiss National Fund SNF, and the Spanish MICINN is 
gratefully acknowledged. This work was also supported by 
the Marie Curie program, by the CPAN CSD2007-00042 and MultiDark
CSD2009-00064 projects of the Spanish Consolider-Ingenio 2010
programme, by grant DO02-353 of the Bulgarian NSF, by grant 127740 of 
the Academy of Finland, by the YIP of the Helmholtz Gemeinschaft, 
by the DFG Cluster of Excellence ``Origin and Structure of the 
Universe'', by the DFG Collaborative Research Centers SFB823/C4 and SFB876/C3,
and by the Polish MNiSzW grant 745/N-HESS-MAGIC/2010/0.
%
%


\clearpage

\end{document}